\documentclass[twocolumn,preprintnumbers,amsmath,amssymb]{revtex4-1}
\usepackage{graphicx}
\usepackage{dcolumn}
\usepackage{bm}
\usepackage{color}

\begin{document}
\title{Thermal Properties of Vortices on Curved Surfaces}

\author{Leopoldo R. G\'omez$^{1}$}
\email{lgomez@uns.edu.ar}
\author{Nicol\'as A. Garc\'ia$^{2}$}
\author{Daniel A. Vega$^{1}$}
\author{Jos\'e Lorenzana$^{3}$}

\affiliation{$^1$ Instituto de F\'isica del Sur
(IFISUR),Departamento de F\'isica, Universidad Nacional del Sur (UNS), CONICET,
Av. L. N. Alem 1253, B8000CPB - Bah\'ia Blanca, Argentina.\\
$^2$ Institut Laue-Langevin, 71 Avenue des Martyrs, 3842 Grenoble, France. \\ $^3$ ISC-CNR, Dipartimento di Fisica, Universit\`a di Roma ``La Sapienza'', Piazzale Aldo Moro 2, 00185 Roma, Italy.}

\date{\today}

\begin{abstract}
We use Monte Carlo simulations to study the finite temperature behavior of vortices in the XY-model for tangent vector order on curved backgrounds. Contrary to naive expectations, we show that the underlying geometry does not affect the proliferation of vortices with temperature respect to what is observed on a flat surface. Long-range order in these systems is analyzed by using the classical two-point correlation functions. As expected, in the case of slightly curved substrates these correlations behave similarly to the plane. However, for high curvatures, the presence of geometry-induced unbounded vortices at low temperatures produces the rapid decay of correlations and an apparent lack of long-range order.
Our results shed light on the finite-temperature physics of soft-matter systems and anisotropic magnets deposited on curved substrates.
\end{abstract}

\maketitle

\section{Introduction}
After more than thirty years of intense research, today it is well-established that several two-dimensional condensed-matter systems, which break a continuous symmetry, disorder with temperature following the Kosterlitz-Thouless (KT) transition \cite{ChaikinLubensky,JoseBook}.
In this transition topological defects, such as vortices in superfluids or superconductors, or disclinations in crystals or liquid-crystals, play a fundamental role. Here, while at low temperatures vortices and anti-vortices are tightly bound in dipoles, at a critical temperature a  topological phase transition occurs, leading to the unbinding of vortices and the disordering of the phase.

The general features of the KT transition are most clearly revealed through the XY model, which in its continuum version is described by the energy \cite{ChaikinLubensky}:
\begin{equation}
\label{eq:enflat}
F=\frac{K}{2} \int d^2\textbf{r} [\nabla \theta]^2,
\end{equation}
where $\theta(\textbf{r})$ is an angle-valued field with values varying from $0$ to $2\pi$, and $K$ is a stiffness associated with the energy cost of inhomogeneities in $\theta$. This simple model describes equally well magnetic systems or liquid crystals, where $\theta(\textbf{r})$ represents local orientations of spins or molecules, or quantum systems like superfluids or superconductors where  $\theta(\textbf{r})$ represents the phase of a collective wave function.

The XY model is known to have two kind of excitations, which at low temperatures are rather independent. From one side there are smooth variations of $\theta(\textbf{r})$, the spin waves, which destroy long-range order at low temperatures.
The other excitations are the vortices, which are point-like singularities. Here the change of $\theta$  in a closed path surrounding a vortex satisfies $\oint \nabla \theta \cdot dl =q$, where $q$ is the charge of the vortex.

The energetic contribution of a configuration with $N$ vortices in a flat surface takes the form \cite{ChaikinLubensky}:
\begin{equation}
\label{HamiltonianVorticesPlane}
\frac{H}{K}=- \sum_{i < j} q_i q_j \ln \frac{\mid \textbf{r}_i - \textbf{r}_j \mid}a+E_c \sum_i q_i^2
\end{equation}
Here, the singular nature of the excitations requires the introduction of a short distant cutoff $a_0$, namely the vortex core radius, and
the core energy $E_c$ (in units of $K$) which takes into account short range energetic contribution beyond the continuum description of Eq.~\eqref{eq:enflat}. Since vortices interact like two-dimensional Coulomb charges Eq.~\eqref{HamiltonianVorticesPlane} is known as the Coulomb gas model \cite{Minnhagen}.

A renormalization-group analysis of the Coulomb gas demonstrated the existence of the KT transition. Here while at low temperatures
the phase is characterized by power law decaying correlations and bound vortex-antivortex pairs, above a critical temperature $T_c$ the unbinding of vortices leads to the disordering of the phase and exponentially decaying correlations \cite{KosterlitzThouless1973,Kosterlitz1974}.
This disordering scenario have been found to describe a huge variety of
two-dimensional systems systems like anisotropic magnets \cite{vaz2008}, superfluids, superconductors and several soft condensed-matter systems like liquids crystals, polymers, colloids,  and others \cite{Strandburg}.

Much less is known about the disordering mechanism and the fate of the KT transition in two dimensional systems which are not flat, but have some degree of curvature. This line of research started by considering the properties and KT transition of Helium in packed powders \cite{KotsuboWilliams1984}, which derived in the study of the properties of the XY model in non-Euclidean geometries \cite{KotsuboWilliams1984}-\cite{Dasmahapatra}.

Such early works, related to quantum condensed phases like superfluids and superconductors, showed that the geometry may modify the KT transition in unexpected ways. In this sense, while the KT transition was found to remain almost unmodified on the surface of spheres \cite{KotsuboWilliams1984,Ovrut}, on surfaces of constant negative curvature (pseudospheres) the critical temperature was found to shrink to zero. 
This means that underlying constant negative curvature strongly affects the KT transition, disordering the system at any finite temperature \cite{CallanWilczek,Dasmahapatra}.

Similarly, it has been suggested that the underlying geometry could also affect the main features of the KT transition for systems related to crystals or liquid-crystal phases \cite{VitelliPRE,Selinger,Brito} on non-Euclidean geometries. Also, recent work on first-order phase transitions on curved geometries have shown that the dynamics of nucleation and growth can be strongly modified by the underlying geometry \cite{Meng,GomezNatComm}.

In this work we use a modified XY model and Monte Carlo simulations to study the temperature behavior and KT transition of tangent vector order on curved surfaces. Given the observed opposite effects of positive and negative constant curvature \cite{KotsuboWilliams1984,Machta,Ovrut,CallanWilczek,Dasmahapatra}, here we focus our study on surfaces shaped as Gaussian bumps. These surfaces are interesting because they have both, positively and negatively curved regions, have the topology of the plane, and can be obtained in the laboratory by relaxing corrugated substrates \cite{HexemerThesis,VegaSoftmatter}.

This paper is organized as follows: In section \ref{Sec:Model} we present details on the model, the simulations, and the geometry used. In section \ref{Sec:Results} we show the main results of these work, regarding the temperature behavior of vortices and short and long-range correlations in these non-Euclidean systems. In section \ref{Sec:Discussion} we discuss and argue on the main reasons which lead to a proliferation of vortices independent on the underlying geometry, as presented in the results. Finally, in section \ref{Sec:Conclusions} we present the main conclusions of this study.

\section{Model and Simulations}
\label{Sec:Model}
Condensed systems formed by spins or stiff molecules constrained to be tangent to a curved substrate can be specified by a unit vector field of the form $\textbf{m}=\cos[\theta(\textbf{r})]\textbf{e}_1+\sin[\theta(\textbf{r})]\textbf{e}_2$, where $\textbf{e}_\beta$ ($\beta=1,2$) are the orthonormal tangent-plane basis vectors \cite{MacKintosh}.
On arbitrary curved geometries, the XY model related to vector tangent order in the continuum can be written in the form \cite{VitelliTurner,BowickAdvPhys,TurnerRMP}:
\begin{equation}
\label{freeenergyfunctionalVector} F=\frac{K}{2}\int  d^2\textbf{r} \, \sqrt{g} \, g^{\beta \gamma} \, [\partial_\beta
\theta(\textbf{r}) - \Omega_\beta(\textbf{r})] \, [\partial_\gamma \, \theta(\textbf{r})- \Omega_\gamma(\textbf{r})]
\end{equation}
Here, points on the surface are specified by a system of curvilinear coordinates $\textbf{r}=(x_1,x_2)$, such that an infinitesimal arc length $ds$ is given in Einstein notation by $ds^2\equiv|d\textbf{r}|^2=g_{\beta\gamma}dx^\beta dx^\gamma $, where $g_{\beta\gamma}$ is the metric tensor, and $\sqrt{g}$ is its determinant. The field $\Omega_\beta(\textbf{r})$ is a connection that compensates for the rotation of the 2D basis vectors $\textbf{e}_\beta$ with respect
to which $\theta(\textbf{r})$ is measured. The connection is intrinsically related to how curved is the surface, such that its curl is equal to the Gaussian curvature of the surface $G(\textbf{r})$ \cite{VitelliPRE}.

In this XY model the connection field is necessary in order to represent the frustration imposed by the geometry. Spins or molecules constrained to be tangent to a curved substrate cannot be all oriented parallel to their neighbors, such that orientational order is geometrically frustrated \cite{Grason}. This is completely different to Euclidean systems
[Eqn.~\eqref{eq:enflat}] or non-Euclidean systems related to quantum collective phases, where there is no frustration and the minimum energy configuration is always given by an homogeneous configuration of $\theta$ \cite{VitelliTurner,BowickAdvPhys,TurnerRMP}.

\begin{center}
\begin{figure}[b]
\includegraphics[width=8.0 cm]{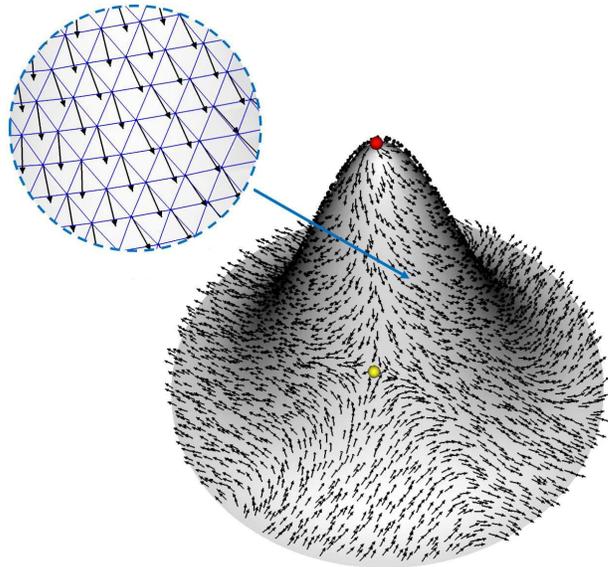}
\caption{Low temperature configuration ($k_{B}T=0.1J$) of the modified XY model obtained by Monte Carlo simulations on a Gaussian bump surface. In this relaxed configuration a positive vortex (red circle) locates on the top of the bump, and a negative vortex (yellow circle) down the surface. The inset shows  a detail of the curved grid used in the simulations and the tangent vector field on this grid.}
\end{figure}
\end{center}

The energy of this non-Euclidean XY model can also be decomposed into a regular spin-wave contribution and a Coulomb gas. However, here the energy of vortices not only includes the vortex-vortex interactions and core energies as in Eqn. 2, but also an interaction with the substrate through a geometric energy \cite{VitelliTurner,BowickAdvPhys,TurnerRMP}:
\begin{equation}
\label{HamiltonianVortices} \frac{H}{K}=\sum_{i<j} q_i \, q_j V(\textbf{r}_i,\textbf{r}_j)+{E_c} \sum_i q_i^2+\sum_i E_i(\textbf{r}_i),
\end{equation}
where $V$ is the interaction between vortices on the curved surface, obtained by the equation  $\Delta_{LB} V(\textbf{r}_i,\textbf{r}_j) = -\delta(\textbf{r}_i,\textbf{r}_j)$, where the Laplace-Beltrami operator is given by $\Delta_{LB}= \frac{1}{\sqrt{g}}\frac{\partial}{\partial
x^\beta}(g^{\beta \gamma}\,\sqrt{g} \, \frac{\partial}{\partial x^\gamma})$, and $E_i(\textbf{r}_i)$ represents the energetic interaction between a vortex and the substrate's topography:
\begin{equation}
\label{eqn:energyvortices}
E_i(\textbf{r}_i)=\left(q_i-\frac{q_i^2}{4\,\pi}\right) \, U_G(\textbf{r}_i)
\end{equation}
Here, $U_G(\textbf{r})$ is the geometric potential which is fixed by the substrate's curvature through the Poisson-like equation $\Delta_{LB} U_G(\textbf{r}) = G(\textbf{r})$.
This interaction implies that when a vortex is placed on a curved surface, it feels a force as if there were a background topological charge proportional to the Gaussian curvature of the substrate\cite{VitelliTurner,BowickAdvPhys,TurnerRMP}. As a consequence of this purely geometric interaction, in general positive (negative) vortices tend to be attracted to regions of local positive (negative) Gaussian curvature. It is interesting to note that due to the asymmetry in the prefactor of Eqn. 5, in general positive and negative vortices have different energies. For example, for $2\pi$ vortices, the prefactor of $U_G(\textbf{r}_i)$ in Eqn. 5 gives a prefactor of $\pi$ for positive vortices and $-3\pi$ for negative vortices.

In order to develop the Monte Carlo simulations which allow to study the  degree of order in these systems as a function of temperature, we first generated homogeneous curved meshes by a combined fast marching - node interaction approach. Here, an initial (inhomogeneous) grid is first obtained by the fast marching method \cite{Sethian}, and it is later relaxed by allowing the nodes to interact with their neighbors with an harmonic potential \cite{Shimada}. This approach leads to homogeneous grids in arbitrary geometries consisting of a triangular tessellation of the surface (see inset in Fig. 1) .

We then numerically study the features of vector tangent order on surfaces by locating unitary vectors on the grids points, such that these vectors are restricted to the local tangent plane of the surface (see Fig. 1), and relaxing their configurations at a fixed temperature by the standard Monte Carlo Metropolis algorithm. Instead of using the continuum model Eqn. 3 we use a simpler discrete Hamilatonian approach. In order to reproduce the geometric frustration of tangent vector order, we need to compare the orientation of two neighboring vectors at the surface. Here, it is necessary to perform the parallel transport of one of the vectors to the position of the other. To do this we follow the numerical approach proposed by Ramakrishnan, Kumar, and  Ipsen, where the Hamiltonian of the system is written as \cite{Ipsen}:
\begin{equation}
\label{HamiltonianIntrinseco} H_{vector}=-\sum_{<i,j>} J \cos \theta_{ij},
\end{equation}
where $\cos \theta_{ij}=\textbf{m}_j \cdot \Gamma_{ij} \textbf{m}_i$. Here, $\textbf{m}_i$, $\textbf{m}_j$ are vectors at neighbor nodes $i$ and $j$, and the operator $\Gamma_{ij}$ brings (parallel transported) $\textbf{m}_i$ into the tangent plane of the
vertex $j$:
\begin{eqnarray}
\Gamma_{ij} \textbf{m}_i&=&[\textbf{m}_i \cdot \textbf{t}_i] \textbf{t}_j+\\ \nonumber
&& \{\textbf{m}_i \cdot [\textbf{N}_i \times \textbf{t}_i]\} \{\textbf{N}_j \times \textbf{t}_j\} \nonumber
\end{eqnarray}
where $\textbf{N}_i$ is the normal at vertex $i$ and $\textbf{t}_i$ is the projection of the unit vector connecting vertex $i$ to its neighbor $j$, to the tangent plane at the $i$ vertex.
Note that at zero temperature $J \equiv K$ in Eqn. 3. However at finite temperatures one needs to renormalize $K$ to take into account temperature dependent effects of spin-wave excitations which otherwise would be neglected in the Coulomb gas model  Eqn. 4 \cite{David}.

For the sake of concreteness, here we focus in studying the properties of the XY model Eqn. 6 on Gaussian bump surfaces, such as shown in Fig. 1. These surfaces have Monge parametrization of the form:
\begin{equation}
\textbf{R}(r,\phi)=r \cos(\phi) \, \textbf{i}+r \sin(\phi) \, \textbf{j}+\alpha r_0 \exp(-r^2/2r_0^2) \, \textbf{k}
\end{equation}
Here, $r,\phi$ are the plane polar coordinates, $\{\textbf{i},\textbf{j},\textbf{k}\}$ is the Euclidean base, and $\alpha$ is a parameter related to the aspect ratio of the bump (for higher values of $\alpha$ the bump is more pronounced). As in other surfaces of revolution, \emph{meridian} and \emph{parallel} curves are defined by setting the polar variables to constant values (along meridians $\textbf{R}=\textbf{R}(r)$, and along parallels  $\textbf{R}=\textbf{R}(\phi)$)

These surfaces are interesting because they have the topology of the plane but a variable Gaussian curvature given by $G(r)=\frac{\alpha^2 e^{-r^2/r_0^2}}{r_0^2 l(r)^2}(1-\frac{r^2}{r_0^2})$, where $l(r)=1+\alpha^2 \, r^2 \, \exp(-r^2/r_0^2)/r_0^2$. The geometric potential associated to this surface takes the form  $U_G(r)=-\int_r^\infty \frac{\sqrt{l(r')}-1}{r'} \, dr'$, which for high-enough values of $\alpha$ tends to attract positive vortices to the top of the bump, and locate negative defects in the region of negative curvature for $r \geq r_0$ \cite{VitelliPRE}.

For the Monte Carlo simulations we started from a random configuration and equilibrated the system at the highest temperature studied ($k_B T=4 J$) and then reduced the temperature in steps of $0.04 J$. At each temperature we allow the system to relax by performing 150000 passes (after one pass all spins of the system have been updated by Metropolis).
We checked that this protocol allows to reach thermal equilibrium and all thermodynamic quantities become stable. Quantities of interest in this study, like correlations and vortex densities, are obtained by averaging over $100$  runs starting from independent random states.

We fix the units of length so that the distance between nodes is close to $a=1$.
The Gaussian bump is defined in a squared domain of size $128\times 128$ and we fix the parameter of the Gaussian $r_0=10$.
The number of nodes results to be around 20000.
We use \emph{open} boundary conditions. Lattice points in the border are treated in the same form as other nodes, but the only difference is that they typically have less neighbors. Periodic boundary condition could also be used, but they can distort the geometry and can affect the number and locating of defect at low temperatures.

The number and location of positive and negative vortices at any temperature is obtained by discretizing the integral $n=-1/2\pi \oint \nabla \theta \cdot dl$ around a closed loop on each elementary triangle plaquette \cite{Phillips}. In the case the closed path encircles a vortex, $n$ takes a nonzero integer value (usually $\pm1$), and the sign of $n$ indicates the chirality of the vortex with respect to the face normal to the surface.

Figure 1 shows a low temperature configuration of this XY model on a Gaussian bump substrate for $\alpha=5$ and $r_0=10$. Note the presence of the unbounded vortex dipole as a consequence of the large vortex-curvature interactions mediated through the geometric potential, which tend to locate a positive vortex on top of the bump, and a negative vortex in the region $r \sim r_0$ \cite{VitelliPRE}.

\begin{center}
\begin{figure}[t]
\includegraphics[width=8.5 cm]{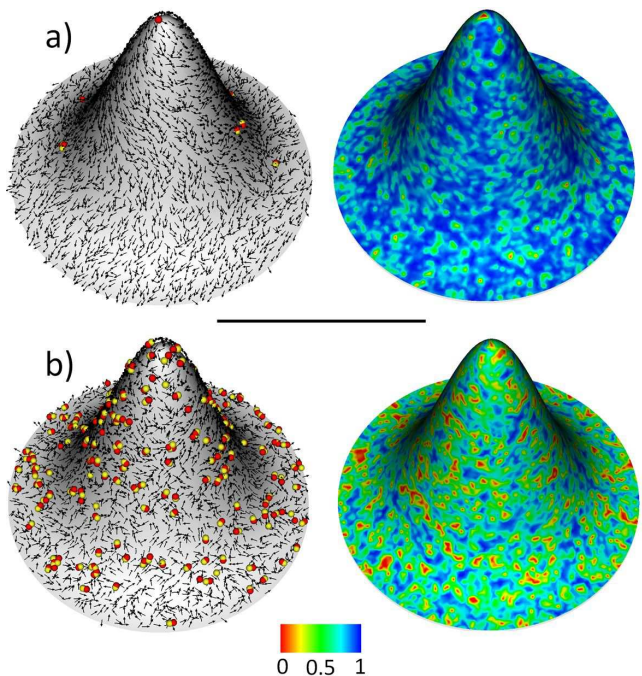}
\caption{Typical vector field configurations obtained by Monte Carlo simulations on a substrate of $\alpha=3, $for low $k_B T=1.0J$ (a), and high $k_B T=1.75J$ (b) temperatures. The left panels show the vector field and the vortices, and the right panels show the corresponding short-range order  maps and the color code used.}
\end{figure}
\end{center}

\section{Results}
\label{Sec:Results}

\subsection{Thermal Properties of Vortices and short-ranged correlations}
For surfaces of varying positive and negative curvature, in principle one may speculate on some possible effects of the non-Euclidean geometry on the KT transition. First, the varying geometry could modify the value of critical temperature $T_c$ where the system disorders. Another possibility is that
curvature acts as a correlated potential leading to a broadening of the transition \cite{maccari2017}.
Here, in order to study the role of curvature on the KT transition, we perform Monte Carlo simulations at different temperatures, for various aspect ratios $\alpha$ of the Gaussian surfaces.

\begin{center}
\begin{figure}[b]
\includegraphics[width=8.0 cm]{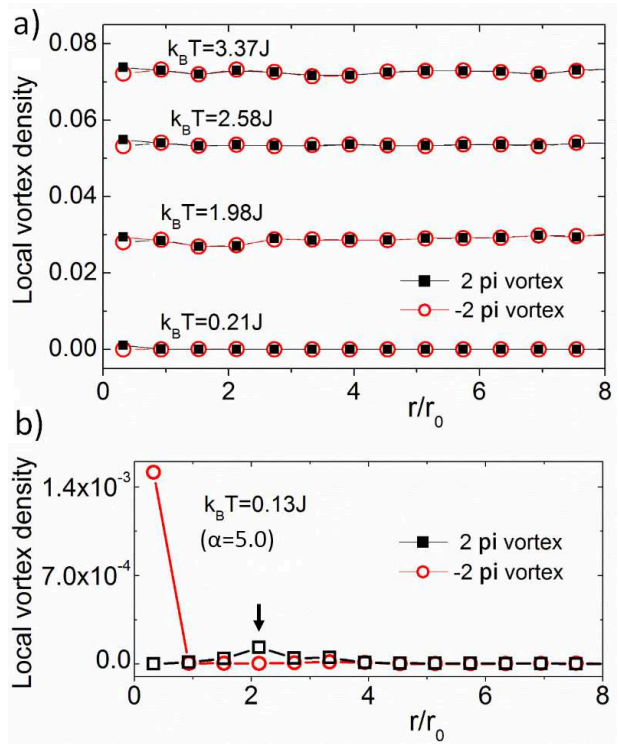}
\caption{a) Local vortex density as a function to the distance $r$ to the centre of the surface, for a substrate of $\alpha=3$ and different temperatures (error bars are of the order of symbol size and omitted for clarity). b) Local vortex density as a function of $r$ at a low temperature,  for a substrate of higher curvature with $\alpha=5$. In this case, there is a gemetrically-unbounded dipole at low temperatures, such that the density of positive and negative vortices is inhomogeneous. These local density curves show that the positive and negative vortices locate at $r\sim 0$ and $r \sim 2r_0$ (see arrow), as expected. As temperature is increase, more dipoles are exited, and the density of positive and negative vortices become homogeneous as in panel a).}
\end{figure}
\end{center}

Figure 2 shows typical configurations of the XY model on a Gaussian bump of a medium aspect ratio $\alpha=3$, for low $k_B T=1.0J$ (a), and high $k_B T=1.75J$ (b) temperatures.
On the left we show the vector field configurations with the positive (red) and negative (yellow) vortices.  As a first rough diagnostic tool of inhomogeneities, on the right we show a map of the short range correlations defined as $S_i=\frac{1}{N_{neigh}}\sum_j \textbf{m}_i \cdot \textbf{m}_j$, obtained by averaging the orientation of a vector with its first neighbors defined in such a way that  $S_i\equiv 1$ signals perfect local order.   At low temperatures, Fig. 2a shows an isolated positive vortex near the top of the bump, a few thermally exited dipoles, and small variations in the vector orientations, which corresponds to the spin waves.
Here the local order map fluctuates around the perfect order ($S_i \sim 1$). As temperature increases, Fig. 2b shows the appearance of more vortices in the form of dipoles, and an increasing disorder which is clearly seen from the short-range order map, which now fluctuates around smaller values of $S_i$ ($S_i \sim 0.5$).

Note that qualitatively, the disordering process seems to be similar to that observed in two dimensional flat systems, and in addition, it seems that the whole process is rather homogeneous (at a given temperature the degree of disorder is similar, independent of the underlying local curvature).

In order to address this more quantitatively, in Fig.~3a) we show the temperature behavior of the local density of positive and negative $2\pi$-vortices, averaged in the azimuthal direction, as a function to the distance to the centre of the surface $r$ (we average on vortices located at a distance between $r$ and $r+dr$ from the top of the bump), at different temperatures. The homogeneity in the disordering process is evident from this plot. For any temperature the density of defects is the same, independently of the region of location on the curved substrate. Note also that the densities of positive and negative defects are the same, as energetically expected for a substrate with the topology of the plane and free boundary conditions \cite{VitelliPRE}. An analogous behavior is observed for the azimuthal-averaged short range order correlator $S_i$, as a function of $r$.

\begin{center}
\begin{figure}[t]
\includegraphics[width=8.5 cm]{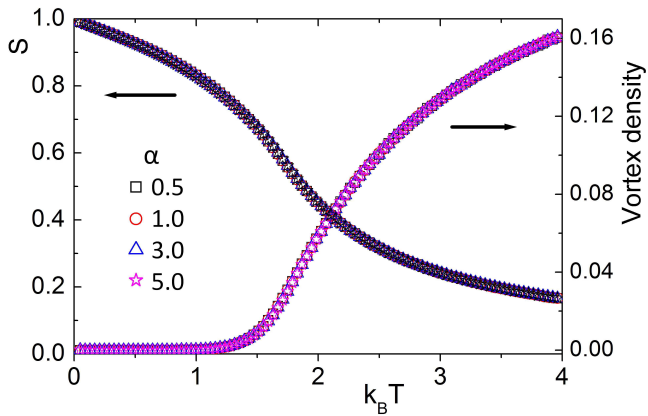}
\caption{Mean local order correlator $S$ and vortex density $\rho$ (obtained by averaging over the whole substrate) as a function of temperature $T$, and for surfaces of different aspect ratios $\alpha$. Note that all the curves superimpose showing that the systems disorder in an equal form, irrespective of the underlying curvature.}
\end{figure}
\end{center}

It is interesting to note here that for substrates of high curvature the low-temperature vortex densities are not homogeneous. In such cases a geometrically-unbounded dipole, such as shown in Figure 1, is found at low temperatures. This produces inhomogeneous local vortex densities, as shown in Fig. 3b, with a peak at $r\sim0$ for positive vortices, and a peak at $r\sim r_0$ for negative vortices. Note that the peak for the negative vortex is much less pronounced as compared with the positive vortex. This is because negative vortices are less confined in the Gaussian bumps (positive vortices are highly confined to a small region of positive curvature for $r\sim0$). However, as temperature is increased, more dipoles are exited and the vortex densities became more homogeneous, such as shown in Fig. 3a.

Having shown that on a surface the disordering is homogeneous, we now compare the thermodynamic behavior for substrates of different curvature. In Fig. 4 we show the temperature behavior of total density of vortices $\rho$ and mean short-range correlator $S=<S_i>$ as a function of temperature. In this plot $\rho$ and $S$ are obtained by averaging on the whole surface, over a hundred independent configurations. Here the different symbols correspond to substrates of different aspect ratios $\alpha$.
Remarkably, as evident from the figure, the thermodynamic behavior is identical for all the geometries within the numerical error. Notice that no rescaling is needed to achieve this result. From this numerical result we conclude that the disorder proceeds independently of the underlying curvature and we speculate that the critical temperature remains identical to the one of a planar hexagonal lattice, namely  $k_B T_c \sim 1.5J$.

\subsection{Long-ranged correlations in curved geometries}

In an Euclidian system the long-range two-point correlation function is defined as $C(r)=\langle \textbf{m}(0) \cdot \textbf{m}(r) \rangle$ \cite{ChaikinLubensky}. As discussed in the introduction, long-range correlations are key to describe the KT transition in planar geometry.

\begin{center}
\begin{figure}[t]
\includegraphics[width=7.5 cm]{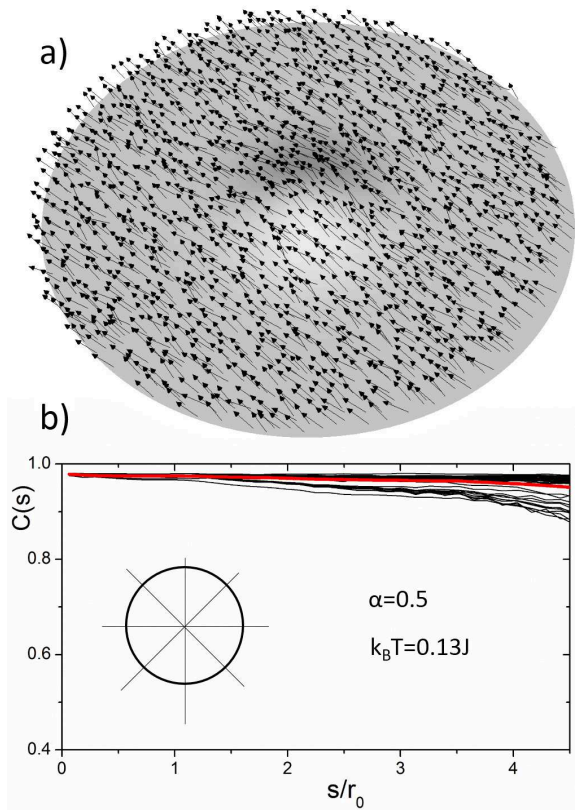}
\caption{Behavior of long-range correlation functions at a low temperature $k_B T=0.13J$, for a system of low curvature $\alpha=0.5$. a) Relaxed spin configuration at this temperature. Note that spins align along a particular direction and there are no vortices in the pattern. b) Two-point correlation function $C(s)$ as a function of geodesic distance, calculated along different meridians (black lines), and the average (red line). Here the correlation $C(r)$ accurately captures the long-range order of the system as observed in panel a), where $C(s) \sim 0.9$ even at long distances. The inset is an scheme showing the different meridians and the characteristic circle $r=r_0$ of the Gaussian surface. }
\end{figure}
\end{center}

On a curved geometry, in order to correctly take the inner product between two distant spins, one of the spins has to be parallel transported to the location (lattice point) of the other, and in principle the correlation function depends on the path chosen to make the parallel transport \cite{RubinsteinNelson}. For the sake of simplicity, here we only calculate correlations between spins which are located in the same meridians ($\phi \equiv 0$ in the Monge parametrization Eqn. 8). In such cases the calculation of correlations are simpler because, due to the azimuthal symmetry, vectors do not rotate as parallel transported along meridians. As  will be clear below, the behavior of correlation functions changes dramatically as a function of curvature. The reason for this change is  more easily visualized if correlations are displayed for a single snapshot of the Monte Carlo simulation. Therefore, below we present single snapshot results computed after thermalization, but without averaging over the initial realizations.

Figure 5a shows a low temperature ($k_B T=0.13 J$) configuration of the XY model on a slightly curved Gaussian with $\alpha=0.5$. In this low-curvature substrate, at low temperatures the spins are well aligned along an arbitrary direction, and the structures do not show unbounded vortices. Figure 5b shows the behavior of the the long-range correlation $C(s)$ as a function of the geodesic distance $s$ between spins, calculated along different meridians (black lines), and an average on the different meridians (red line).  In this case the two-point correlation function behaves very similar to what is obtained in the plane, showing that the system is well ordered, with $C(s)$ almost constant ($C \sim 0.95$) or decaying very slowly. Note also the small dispersion of the correlation calculated along the different meridians.

On the contrary, Fig. 6a shows a completely different behavior of two-point correlation for a spin configuration on a highly curved substrate with $\alpha = 7$, at the same low temperature $k_B T=0.13 J$. Here along some meridians the correlation function decay slowly (pink line in Fig. 6a), but along other meridians the correlation decay abruptly for distances of the order $r \sim 2 r_0$ (blue line in Fig. 6a).

The huge dispersion in the correlation functions obtained along different meridians shown in Fig. 6a is mainly due to the presence of a geometry-induced unbonded dipole. Here due to the high curvature of the substrate, a positive vortex locates on the top of the Gaussian and a negative vortex locates around $r \sim 2 r_0$ (this unbounded dipole exist even at $T=0$, as pointed out before). Panels 6a and 6b show two views of a snapshot of the spin configuration at this temperature, with the positive (negative) vortex indicated with a red (yellow) sphere.

Indeed, the presence of this unbounded dipole completely distorts the behavior of the two-point correlation function. There are some paths, such as the as the meridian indicated with the dashed line in Fig. 6b, where the orientations of spins change slightly, leading to a slowly decaying $C(s)$ (the correlation function along this path is plotted in Fig. 6a with the pink line). But for correlations calculated along meridians which are in the neighborhood of the negative vortex, such as the path indicated with a dashed blue line in Fig. 6c, the orientations of spin change as a consequence of the unbounded vortex, producing an abrupt decay in the correlation function (the correlation along this path is plotted in Fig. 6a with the blue line).  Note that if one averages over initial realizations the azimuthal symmetry is recovered as the azimuthal position of the negative vortex is arbitrary.

It is clear that the system is still strongly correlated in this high curvature case, but the simple two-point correlation function does not reflect such degree of order. Detecting the order in these cases may require the computation of three-point correlations functions, so that paths are restricted to specific azimuthal distances to the unbound negative vortex. Thus, in these and other non-Euclidean systems, long-range order may be much harder to characterize, making difficult the comparison with the results obtained in the plane.

\begin{center}
\begin{figure*}[t]
\includegraphics[width=12 cm]{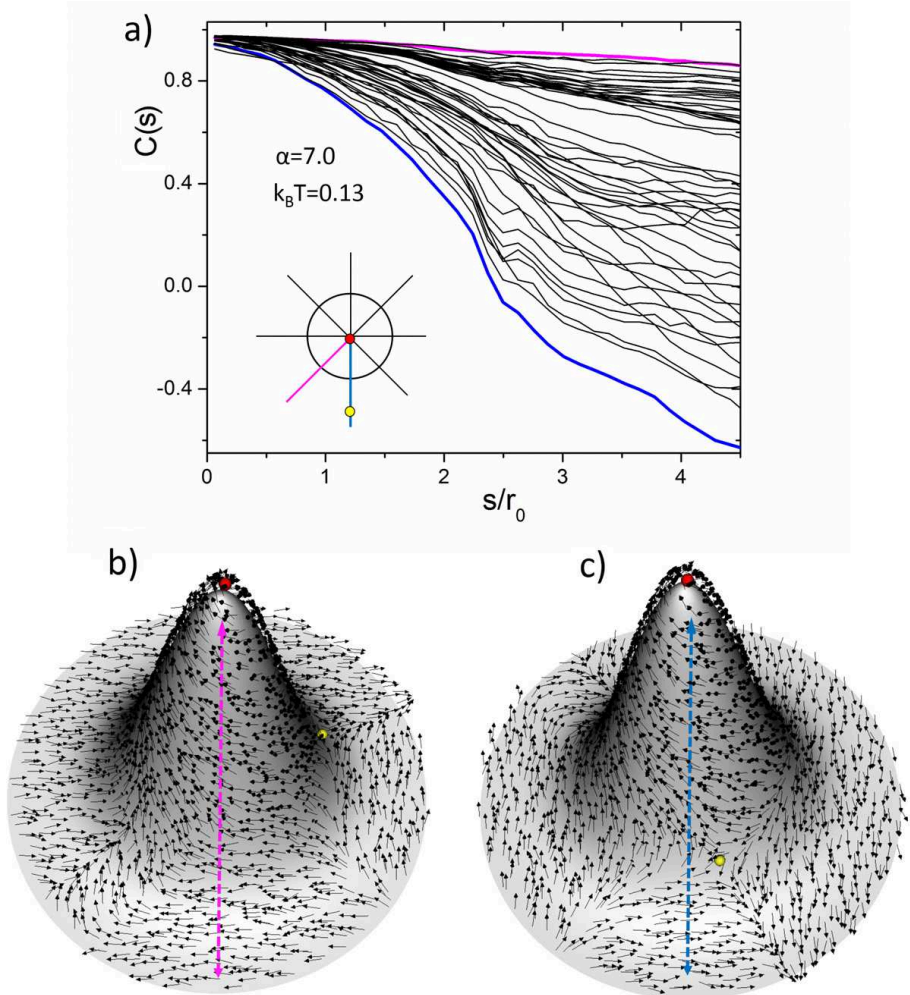}
\caption{Behavior of long-range correlations at a low temperature $k_B T=0.13J$, for a highly curved system with $\alpha=7.0$. a) Two-point correlation function $C(s)$ as a function of geodesic distance $s$, calculated along different meridians (black lines) of the Gaussian bump. Color lines corresponds to correlations calculated along specific meridians, which are indicated with a dashed line in panels b) and c). The inset is an scheme illustrating the different meridian paths used to get the correlations and the location of the vortices. Notice the huge dispersion of correlations calculated along different meridians. b), c) shows to views of the configuration of spins at this temperature. The dashed pink line of panel b) indicates a meridian where the spin orientation slightly changes and two-point correlations depicts long-range order (the correlation calculated along this path is shown in panel a) with the pink line). In panel c) we show a meridian in dashed blue where the orientations of spins changes due to the negative vortex (shown as the yellow sphere). Along this path the two-point correlation decays abruptly (shown in panel a) as a blue line). Thus, in non-Euclidean systems the two-point correlation function may not correctly capture long-range ordered configurations.}
\end{figure*}
\end{center}

\section{Discussion}
\label{Sec:Discussion}

The fact that the geometric potential introduces a highly inhomogeneous energetic landscape for a single vortex \cite{VitelliPRE,VitelliTurner,VitelliTurner,TurnerRMP} [Eqs.~\eqref{HamiltonianVortices},~\eqref{eqn:energyvortices}] appears in strong contradiction with the completely homogeneous and independent on the underlying geometry disordering process. Notice that we have considered a geometry of varying positive and negative curvature, where vortices are attracted (repelled) to regions of same (different) curvature sign, and even further, positive and negative vortices have different energies. We find that these variations are by no means negligibly respect to the temperature. For example for $\alpha=5$ the modulus of the geometric potential has a bump form with a width of the same order of the underling surface and a height given by $U_G(0)-U_G(\infty)=-3.55$ (in units of $K\sim J$). Thus, the energetic scale of the geometric potential is larger than the thermal energy considered (taking into account also the charge prefactors in Eqn. 5), and a naive Boltzman factor for particle activation $\sim \exp\{-[E_c+E_i(r)]/k_BT\}$  would yield a strongly inhomogeneous vortex density.

\begin{center}
\begin{figure*}[t]
\includegraphics[width=14.0 cm]{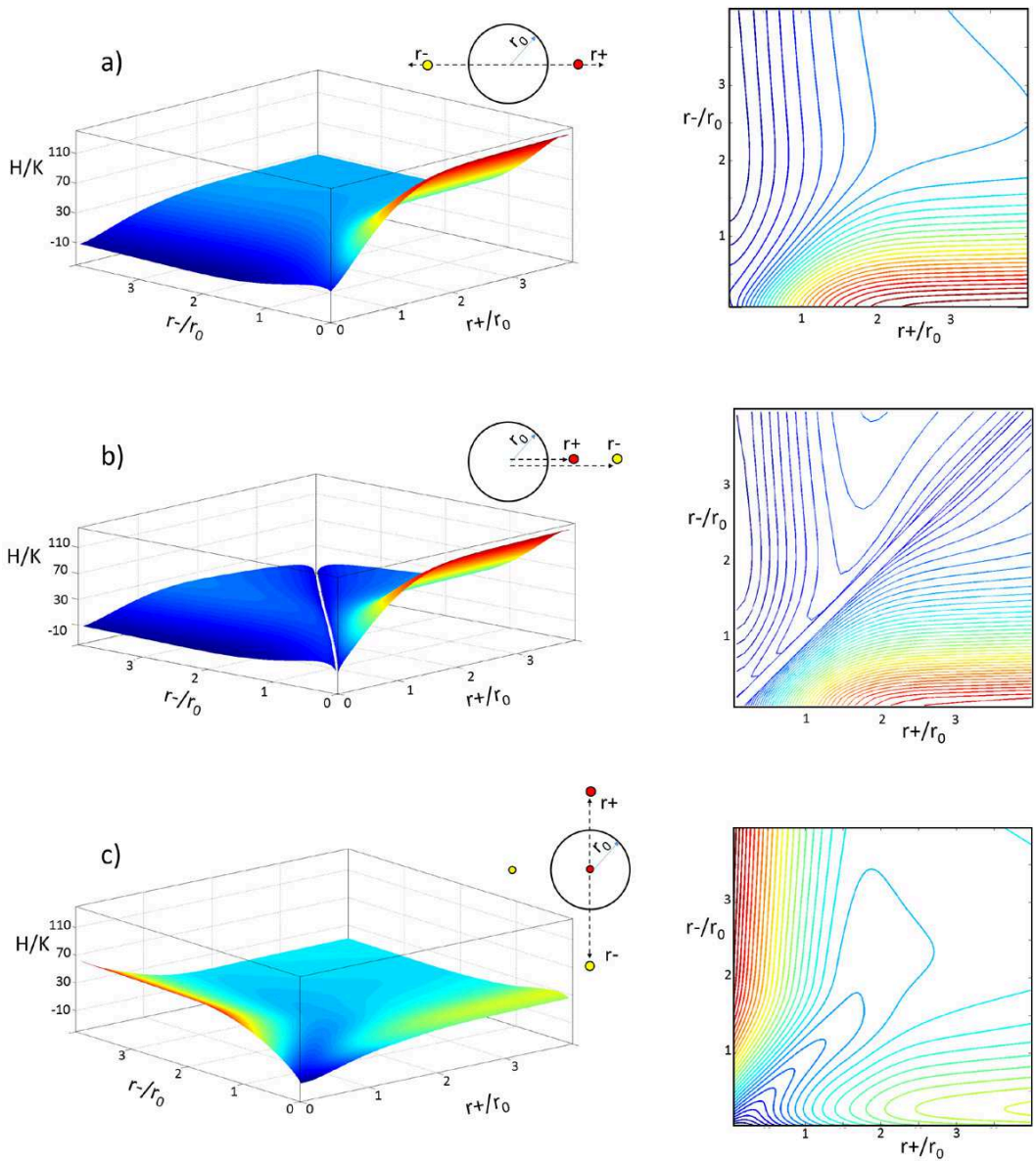}
\caption{Energy landscapes for a dipole and pair of dipoles on a Gaussian bump of $\alpha=7$ (left panels), and the energy level curves (right panels), where we use the same color code as in the left panel to identify the intensities. a) Energy landscape for a first dipole, where the $r_+$ and $r_-$ coordinates corresponds to the position of positive and negative vortex, respectively, in opposite directions to the Gaussian bump (see schematic inset). The energy minima corresponds to $r_+\sim 0$ and $r_- \sim 2r_0$, such that the dipole unbinds due to the geometric force given by the geometric potential.  b) Energy landscape for a first dipole, but now the vortex and antivortex are in the same meridian (see schematic inset). The energy minima still corresponds to $r_+\sim 0$ and $r_- \sim 2r_0$. Note also that the dipole configurations having $r_+ \sim r_-$ have always roughly the same energy, irrespective of the location of the dipole in the surface. c)  Energy landscape for a second dipole, in the same substrate, but when the first dipole is already unbounded (see inset). Now the energy minima corresponds to the second dipole bound, because the geometry has been already largely screened through the unbinding of the first dipole.}
\end{figure*}
\end{center}

To understand this it should be first noted that topological defects are nucleated  as vortex ($q=2\pi$) and antivortex ($q=-2\pi$) pairs. We can estimate the activation energy as the
energy difference between   a  configuration with a pair separated by the microscopic distance $a$ and configuration without pairs. According to Eq.~\eqref{HamiltonianVortices} this is:
\begin{equation}
  \label{eq:activ}
  \frac{E_a}{4\pi^2K}=- V({\bf r},{\bf r}+{\bf a})+2 E_c-\frac{2}{4\pi} U_G({\bf r}),
\end{equation}
where $|{\bf a}|=a$ is a microscopic distance of the order of the lattice node separation. The last term is the contribution of the geometric potential of each vortex and we have neglected differences in $U_G$
at the microscopic distance. The first contribution due to the vortex-antivortex attraction is computed in Ref.~\cite{VitelliPRE},  $V({\bf r},{\bf r}+{\bf a})=-\ln(a^2)/(4\pi)-U_G({\bf r})/(2\pi)$. Therefore, the contribution of the geometric potential cancels and one recovers the result that would be obtained for a flat surface, $E_a/K-2 E_c=-\pi \ln(a^2)$. Our numerical result shows that this estimate of the activation energy is rather robust and the proliferation of vortices results to be independent of the geometric potential. We speculate that this result could change if the core energy would be very large, so that the disordered phase were made of dilute unbound vortices, very far from the configurations considered to derive Eq.~\eqref{eq:activ}.

The above argument does not take into account the physics of tangent vector order, where the connection $\Omega$ arises in the continuous description [Eq.~\eqref{freeenergyfunctionalVector}], and vortex-antivortex pairs can spontaneously unbind even at zero temperature due to the coupling with the curvature. Why this effect does not produce a gas of dilute unbound vortexes at low temperatures?
While the unbinding due to the geometry is very effective of the first pair, this process is inhibited for subsequent pairs due the screening of the preexisting vortices.
To see this more clearly we have used Eq.~\eqref{HamiltonianVortices} to compute the energy at zero temperature for vortices configurations in a Gaussian bump of large curvature. For the vortex-vortex interaction energy $V(\textbf{r},\textbf{r}')$ on the Gaussian bump we have used the expression derived in Ref \cite{VitelliPRE}.

Figure 7a shows the energy for a single vortex-antivortex pair where the vortices are located in opposite meridians of the Gaussian bump (see schematic inset where the positive vortex has $\phi\equiv0$ and the negative vortex $\phi=\pi$). Here $r_-$, $r_+$ represent the distances of the positive and negative vortex to the bump maximum. In the right panel we show the level curves of the energy landscape. As clear from this figure, for a single vortex-antivortex pair the energy is minimized at $(r_+,r_-)\sim (0,2 r_0)$, i.e.  the positive vortex close to the top of the bump, and the negative vortex in the region of negative curvature with $r\sim 2r_0$ (see schematic inset). Thus, in this case the energy is minimized by the unbinding of the vortex dipole, in a configuration similar to the observed in Figures 1 and 6. From Fig. 7a it is also clear that the worst energetic configuration is obtained by locating the negative vortex at the top of the bump, and the positive vortex at the negatively curved region, i.e. $(r_+,r_-)\sim (2 r_0,0)$. Figure 7b shows the energy landscape obtained for a dipole, when the vortex and antivortex are oriented in the same meridian (see schematic inset). Here that the energy minimum is still the unbounded dipole $(r_+,r_-)\sim (0,2 r_0)$. Note also that the bounded dipole obtained for $r_+ \sim r_-$ has roughly the same energy irrespective on the location of the substrate, as discussed in the above paragraphs.

Now consider what happens with a \emph{second} dipole, which can be thought to be exited by temperature when the first dipole has been already unbounded by the geometry. Figure 7c shows the energy landscape for the second dipole, where we consider the first vortex-antivortex pair as fixed. Here $r_-$, $r_+$ represent the distances of the positive and negative vortex of the second pair to the bump maximum (see schematic inset for the vortex configuration), and the right panel shows the level curves. Note that the energy landscape is much flatter for this second dipole, as compared to the first pair, such that the second vortex dipole \emph{sees} a much more homogeneous geometric field than the first dipole. This is because the unbinding of the first dipole has largely screened the substrate geometry/frustration \cite{VitelliPRE}. Thus, new thermally excited dipoles do not feel especial curvature-related forces that tend to unbind them which would manifest in a reduction of $T_c$, which is not the case.  This fact, together with the insensibility to geometric effects when vortex and antivortex are close to each other,  explains the independence of the density of vortices to curvature as shown in Figs. 3 and 4.

\section{Conclusions}
\label{Sec:Conclusions}

Here we have used Monte Carlo simulations on a modified XY model to unveil the role of a non-Euclidean geometry on the Kosterlitz-Thouless transition. Previous studies at $T \equiv 0$ had shown that the energy of vortices is very sensitive to the geometry through a position dependent potential (the geometric potential). Surprisingly, our simulations show that the underlying geometry does not play a role in the disordering of the system, and the thermal properties of vortex are practically indistinguishable from those in 2D flat systems. This is a direct consequence of two effects. First, a dipole nucleation rate independent of the geometry due to the fact that geometric effects tend to cancel when considering neutral pairs at close distances. Second, the low temperature screening of curvature by unbounded vortices, which produce rather homogeneous energy landscapes for new thermally excited dipoles, with no preferential regions for location or extra forces for unbinding.

Regarding long-range order, we have shown that for slightly curved substrates the two point correlation function behaves similarly to the plane, as expected. However, in highly curved substrates, the presence of unbounded vortices at low temperatures produces a wide dispersion of correlations along different paths, such that the two-point correlation function fails to capture long-range correlated configurations.

Because the model we used here is minimal, in the sense that it only has the principal energetic contribution to model tangent order on curved surfaces, the results obtained here should apply to the thermal properties of a variety of soft-matter systems, like crystals or liquid crystals, when restricted to reside in a two-dimensional curved geometry or strongly anisotropic magnetic systems. Possible applications may include a few layers of ferromagnetic Fe deposited on a curved gold substrate\cite{vaz2008}, and block copolymer thin films \cite{HexemerThesis,VegaSoftmatter,GomezPRE}, or liquid crystals \cite{AFernandezNievesReview}, on curved topographies.

\section*{Acknowledgements}
We acknowledge helpful discussions with V. Vitelli and A. M. Turner.
This work was supported by the National Research Council of
Argentina, CONICET, ANPCyT, and Universidad Nacional del Sur.
J.L acknowledges financial support by Italian MIUR under projects
PRIN-RIDEIRON-2012X3YFZ2, and Italian MAECI under collaborative projects SUPERTOP-PGR04879 and AR17MO7.

\end{document}